\documentclass[twocolumn,aps,prl,showpacs,superscriptaddress]{revtex4}  
\usepackage{graphicx}  
\usepackage{graphics}
\usepackage{dcolumn}   
\usepackage{bm}        
\usepackage{amssymb}   
\usepackage{color}

\def\AA{{\cal A}}
\def\BB{{\cal B}}
\def\CC{{\cal C}}

\begin{document}
\title{Exact distributions for stochastic gene expression models with bursting and feedback}
\author{Niraj Kumar}
\affiliation{Department of Physics, University of Massachusetts Boston, Boston MA 02125, USA}
\author{Thierry Platini}
\affiliation{Applied Mathematics Research Center, Coventry University, Coventry, CV1 5FB, England}
\author{Rahul V. Kulkarni}
\affiliation{Department of Physics, University of Massachusetts Boston, Boston MA 02125, USA}
\date{\today}

\begin{abstract}
Stochasticity in gene expression can give rise to fluctuations in
protein levels and lead to phenotypic variation across a population of
genetically identical cells.  Recent experiments indicate that
bursting and feedback mechanisms play important roles in controlling
noise in gene expression and phenotypic variation. A quantitative
understanding of the impact of these factors requires analysis of the
corresponding stochastic models.  However, for stochastic models of
gene expression with feedback and bursting, exact analytical results
for protein distributions have not been obtained so far.  Here, we
analyze a model of gene expression with bursting and feedback
regulation and obtain exact results for the corresponding protein
steady-state distribution.  The results obtained provide new insights
into the role of bursting and feedback in noise regulation and
optimization.  Furthermore, for a specific choice of parameters, the
system studied maps on to a two-state biochemical switch driven by a
bursty input noise source.  The analytical results derived thus
provide quantitative insights into diverse cellular processes
involving noise in gene expression and biochemical switching.
\end{abstract}

\pacs{87.10.Mn, 02.50.r, 82.39.Rt, 87.17.Aa, 45.10.Db}
\maketitle

\section{Introduction}
Cellular responses to environmental fluctuations often involve
biochemical reactions that are intrinsically stochastic. For example,
stochasticity (noise) plays an important role in processes leading to
gene expression~\cite{ouden2008,elowitz10,kondev13} and in biochemical
switching between distinct
states~\cite{koshland1976,berg1977,cluzel11,levin}. Regulation of
noise in these processes is critical for the maintenance of cellular
functions as well as for the generation of phenotypic variability
among clonal cells. Quantitative modeling of mechanisms of noise
regulation is thus a key step towards a fundamental understanding of
cellular functions and variability.

Noise regulation in cells is typically implemented by regulatory
proteins such as transcription factors (TF). Recent research has
demonstrated that, at the single-cell level, regulatory proteins are
often produced in
bursts~\cite{xienature2006,xiescience2006,golding2005,raj2006,singer2006}.
Such proteins can further be involved in autoregulation (e.g. the Tat
regulatory protein which controls the latency switch of HIV-1 viral
infections)~\cite{wein2005,maheshri2010,abhi09,bowsher,
  wang2011,wolen05,martin2008} or in downstream regulation of
biochemical switches (e.g. switching of flagellar rotation states in
bacterial chemotaxis)~\cite{koshland1976,berg1977,cluzel11,levin}.
Some interesting questions arise from these observations: How does
feedback from proteins produced in bursts regulate noise in
gene expression and biochemical switching? How can gene expression
parameters be tuned to optimize noise in the presence of bursting and
feedback? The aim of this Letter is to address these questions by
developing a gene expression model that combines bursting and feedback
for which we obtain the exact stationary distribution.

Previous work on noise in gene expression has focused on exact
analytical solutions for models with: a) bursting but no feedback
effects \cite{swain08} or b) feedback effects but no protein
production in bursts
\cite{wolen05,martin2008,grima,blossey2013}. Similarly, previous work
on noise-induced biochemical switching \cite{levin} does not consider
the case of input noise source produced in bursts. In this letter, we
introduce a \emph{single} model that addresses these gaps in the
field. Our model reduces to multiple previously studied models in
limiting cases. We obtain exact analytical distributions that
significantly extend previously obtained results and lead to new
insights.

\begin{figure}
 \includegraphics[width=8.5cm]{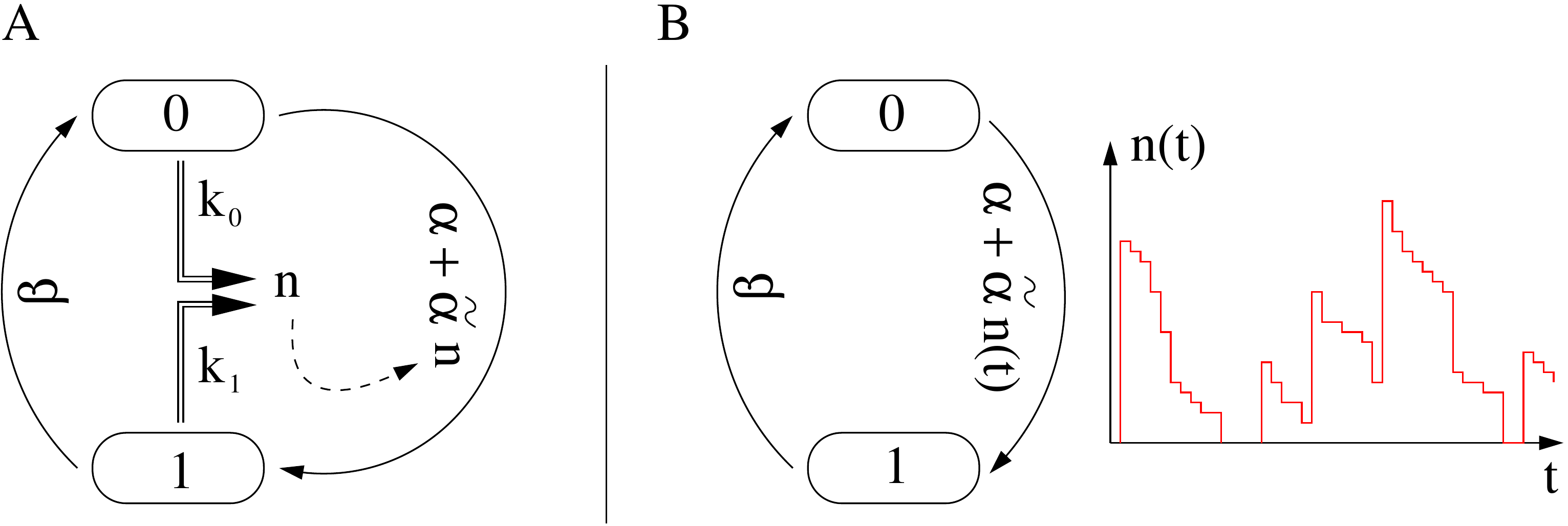}
 \caption{ (A) Schematic representation of the model. Protein bursts
   from inactive(active) state are generated with rate
   $k_0(k_1)$. Rate of transition from inactive to active state is
   $\alpha+\tilde{\alpha}n$, and that from active to inactive is
   $\beta$. (B) For $k_0 = k_1$ the model maps onto a two-state switch
   driven by a bursty input source.}
 \label{fig:model}
 \end{figure}

{\bf Model:} A schematic representation of the model is shown in
Fig.\ref{fig:model}. Here 0 and 1 represent the inactive and active
state of the promoter, respectively.  Note that the terms
inactive/active are simply used to label the two states since protein
production can occur from either state.  Specifically, protein
production from the inactive (active) state occurs with rate $k_0$
($k_1$). Each production event results in a random burst of proteins
and we assume that these bursts are distributed geometrically with
mean size $b$. The degradation rate of proteins is denoted by $\mu$.
The rate of switching from active to inactive state is denoted by
$\beta$. The rate at which the inactive state switches to the active
state has two contributions: the spontaneous contribution with rate
$\alpha$, and the feedback contribution, with rate $\tilde{\alpha}n$
(where $n$ is the number of proteins, and $\tilde{\alpha}$ measures
the strength of the feedback). The linear dependence on $n$ for the
feedback term is consistent with experimental characterization of the
genetic circuit for expression of HIV-1 Tat protein \cite{wein2005}.

Since we allow protein production from both active and inactive
states, the model can be used to analyze the effects of positive
feedback as well as negative feedback. When $k_1>k_0$, the feedback
term enhances protein production leading to positive feedback. In
contrast, $k_1<k_0$ leads to negative feedback. For $k_1=k_0$, protein
production is independent of the promoter state. As indicated in
Fig. 1B, the model then corresponds to a bursty input noise source
controlling switching between the two states. Thus the same model can
be used to analyze the impact of input protein noise on the statistics
of a simple two-state switch \cite{levin}.

Let $P_{\sigma,n}(t)$ denote the probability to find, at time $t$, the
promoter in state $\sigma$ ($\sigma=0,1$) with $n$ proteins in the
cell. The temporal evolution of $P_{\sigma,n}(t)$ is given by the
following Master equations \cite{kampen}:
\begin{eqnarray}\label{MasterEq1}
\partial_t P_{0,n}&=&k_0\sum_{p=0}^ng(p)P_{0,n-p}+\mu (n+1)P_{0,n+1}\nonumber\\&+&\beta P_{1,n}
-[k_0+\alpha+\tilde{\alpha}n+\mu n]P_{0,n},\nonumber\\
\partial_t P_{1,n}&=&k_1\sum_{p=0}^ng(p)P_{1,n-p}+\mu (n+1)P_{1,n+1}\nonumber\\&+&(\alpha+\tilde{\alpha}n) P_{0,n}-[k_1+\beta+\mu n]P_{1,n},
\end{eqnarray}
where $g(n)=b^n/(1+b)^{n+1}$ is the protein burst distribution.
To proceed further, let us define the generating functions
$G_{\sigma}(z,t)=\sum_{n}P_{\sigma,n}(t)z^n$ with
$\sigma=0,1$. Correspondingly, Eq. (\ref{MasterEq1}) can be recast as
\begin{eqnarray}{\label{G0G1}}
\partial_tG_0&=&k_0\tilde{g}G_0+\mu\partial_zG_0+\beta G_1 \nonumber\\
&-& (k_0+\alpha)G_0-(\tilde{\alpha}+\mu)z\partial_zG_0,\nonumber \\
\partial_tG_1&=&k_1\tilde{g}G_1+\mu\partial_zG_1+\alpha G_0+\tilde{\alpha}z\partial_zG_0\nonumber\\&-& (k_1+\beta)G_1-\mu z\partial_zG_1,
\end{eqnarray}
where $\tilde{g}(z)$ is the generating function of the protein burst
distribution given by $\tilde{g}(z) =1/(1+b(1-z))$.  In the long time
limit, Eq. (\ref{G0G1}) is used to derive an equation for the
generating function of the protein steady-state distribution,
$G(z)=G_0(z)+G_1(z)$. After a sequence of transformations (see Supplementary Information) Eq. (\ref{G0G1}) reduces to a 
hypergeometric differential equation, leading to the solution:
\begin{eqnarray}{\label{solution}} 
G(z)&=&\left[\frac{1}{1+b(1-z)}\right]^{k_1/\mu} \\ &&\times \frac{_2F_1[u,v|u+v+1-w|1-\phi\{1+b(1-z)\}]}{_2F_1[u,v|u+v+1-w|1-\phi]},\nonumber
\end{eqnarray}
where the quantities, $u$, $v$, $w$ and $\phi$ are related to model parameters by
$$
u+v=\frac{\Delta k+\alpha+\beta-\tilde{\alpha}k_1/\mu}{\mu+\tilde{\alpha}},~~~~~~
uv=\frac{\beta\Delta k}{\mu(\mu+\tilde{\alpha})},
$$
\begin{equation}
w=\frac{\Delta k+\mu+\tilde{\alpha}(1+b)(1-k_1/\mu)}{\mu+\tilde{\alpha}(1+b)},~~~~
\phi=\frac{\mu+\tilde{\alpha}}{\mu+\tilde{\alpha}+b\tilde{\alpha}},
\end{equation}
with $\Delta k=k_0-k_1$, and $_2F_1$ represents the Gaussian hypergeometric function.
This solution for the generating function is the central result of this
paper.  It can be shown that our result reduces to previously obtained
results in different limiting cases (Supplementary Information).  It
can be used to derive exact analytical results for several quantities
of interest. For example, the steady-state probability that the
promoter is in state $0$ ($P_0= G_0(1)$) is given by (see
Supplementary Information)
\begin{eqnarray}{\label{eP0}}
P_0&=&\frac{\phi\beta}{\alpha+\beta+\frac{k_0\tilde{\alpha}b}{\mu+\tilde{\alpha}(1+b)}}\nonumber\\ &&\times\frac{_2F_1[u+1,v+1,u+v+2-w,1-\phi]}{{_2F_1}[u,v,u+v+1-w,1-\phi]}.
\end{eqnarray}
Furthermore, Eq.(\ref{solution}) can be used to obtain an analytical
expression for the protein steady-state distribution
$P_n=P_{0,n}+P_{1,n}$ and to analyze the corresponding moments. These
expressions lead to quantitative insights into multiple topics of
current research interest as discussed below.

\begin{figure}
 \includegraphics[bb=159 453 437 668,height=6cm,width=9cm]{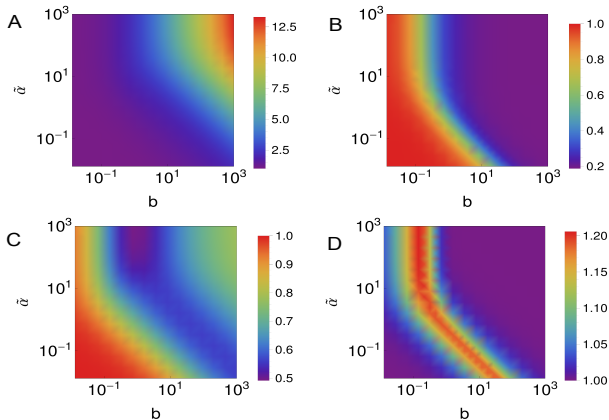}
\caption{Protein noise regulation: In the upper panel, density plots for the
  noise ratio $\eta(\tilde{\alpha})/\eta(\tilde{\alpha}=0)$ as a
  function of $b$ and $\tilde{\alpha}$ for A) negative feedback
  ($k_0=10,k_1=0$) and B) positive feedback ($k_0=0,k_1=10$). In the
  lower panel, comparison between the original and effective models:
  Density plots for $\eta/\eta_{eff}$ are plotted as a
  function of $b$ and $\tilde{\alpha}$ for C) negative feedback
  ($k_0=10,k_1=0$) and D) positive feedback ($k_0=0,k_1=10$). Other
  parameters are: $\alpha=\beta=\mu=1$.}
\label{fig:fig2}
\end{figure}

\noindent{\bf Regulation of protein noise:} There has been
considerable focus in previous work on analyzing the effects of
feedback on the noise $\eta=\langle n^2 \rangle/\langle n \rangle^2-1$
characterizing the protein steady-state distribution
\cite{simpson,wein2005, maheshri2010,abhi09,bowsher}. To analyze the
impact of feedback, we first compare the noise for the case with
feedback ($\tilde{\alpha} > 0$) to the case without feedback
($\tilde{\alpha} = 0$) in Fig.\ref{fig:fig2}(A,B). It is interesting
to observe that negative feedback increases the noise when compared to
the case without feedback: $\eta(\tilde \alpha)/\eta(0)>1$. On the
other hand, positive feedback leads to a decrease of noise
$\eta(\tilde \alpha)/\eta(0)<1$.  While this may appear surprising
given previous results \cite{becskei}, this observation is consistent
with recent results from simulations \cite{bowsher}.  It should be
further noted that negative feedback leads to a reduction in mean
levels whereas positive feedback increases mean levels; thus the
changes in $\eta$ can be driven largely by changes in the mean levels.
It follows that to determine the effects of feedback on noise, it is
desirable to compare models which give rise to the same mean levels.

To address this issue, we introduce an effective model with no
feedback that is characterized by a {\em constant} rate $\alpha_{eff}$
for promoter switching from the inactive to active state. The
parameter $\alpha_{eff}$ is determined analytically (Supplementary Information)
by the condition
that the mean protein levels $\langle n \rangle$ and $P_0$ are
identical in the original and effective models. The remaining
parameters are the same as in the original model.  In the following,
we compare noise in protein distributions for the original and
effective models.

Fig.\ref{fig:fig2}(C,D) illustrates the ratio $\eta/\eta_{eff}$ for
negative as well as positive feedback.  For negative feedback, protein
noise in the original model is lower than the noise in the effective
model, i.e. the effect of negative feedback is to reduce noise.  For
positive feedback, as shown in Fig.\ref{fig:fig2}D, we observe that
feedback increases the noise when compared to the noise for the
effective model.  Thus, in the context of regulation of protein noise,
the results obtained indicate that the choice of reference model plays
a critical role. In relation to the model without feedback, we observe
that negative (positive) feedback increases (decreases) the noise.
However, in relation to the effective model, which preserves the
average number of proteins, the opposite behaviour is observed.

Fig.\ref{fig:fig2}(C,D) also indicates that the effective model
provides a useful approximation to the original model for a wide range
of parameters, in particular for positive feedback. In this case,
for the range of parameters considered in Fig. \ref{fig:fig2}(D), the effective model provides a good 
approximation in regions of parameter space for which a) $\alpha_{\text{eff}}\approx \alpha$ or b) $\alpha_{\text{eff}}\gg\beta$. In the former case, fluctuations in protein levels make a negligible contribution to the promoter switching rate, 
whereas the latter condition represents (almost) constitutive 
production of protein bursts, making the effective model indistinguishable from the 
original model.
However, there is a class of problems for which the
effective model is inadequate and it is necessary to analyze the
complete model.  An important example includes noise optimization in
the presence of feedback by varying system parameters, as discussed in
the following.

\noindent{\bf Noise Minimization:} Recent work \cite{elf} has analyzed
noise minimization due to negative feedback for a model similar to the
one outlined in Fig. 1.  In this model, the binding of a TF switches
its promoter to a repressed state, (i.e. set $k_1 = 0$) and the
switching rate $\beta$ corresponds to the dissociation rate of the TF
from its promoter.  In the limit $\beta\rightarrow\infty$, there is no
feedback, since proteins bursts are effectively produced
constitutively with rate $k_0$. To examine noise minimization, the
system parameters $k_0$ and $\beta$ are varied subject to the
constraint that the mean protein number $\langle n \rangle$ is held
fixed.  In particular, it is of interest to determine: a) the minimum
dissociation rate, $\beta_{\mathrm min}$, required for negative
feedback to result in a reduction of noise relative to the model with
no feedback.  b) the optimal rate $\beta_c$ at which noise suppression
is maximal.  In the following, we explore the insights gained for this
problem (for the model in Fig. 1) using exact analytical results for
moments derived using Eq. (\ref{solution}).

Fig. 3 illustrates the variation of protein noise $\eta$ as a function
of TF dissociation rate $\beta$, keeping the mean protein levels fixed
by changing the transcriptional rate $k_0$. In the limit
$\beta\rightarrow\infty$ (i.e no feedback), we have
$\eta=(1+b)/\langle n \rangle$.  As $\beta$ is reduced, the noise
initially decreases, reaches a minimum value at $\beta = \beta_c$ and
subsequently increases.  In contrast, for the corresponding effective
model with a constant rate of promoter transitions (as defined in the
preceding paragraphs), we have $\eta > (1+b)/\langle n \rangle$ for
all finite $\beta$, i.e. there is no noise reduction.  This indicates
that it is essential to consider the role of fluctuations in the rate
of promoter transitions (from active to repressed state) in
understanding noise reduction due to negative feedback.

The parameter $\beta_{\mathrm min}$ can be determined by the condition
that for $\beta =\beta_{\mathrm{min}}$, we have $\eta = (1+b)/\langle
n \rangle$.  The exact expression for $\eta$ in combination with some
approximations, specifically $P_0 = {\beta}/{(\beta + \alpha +
    \tilde{\alpha} \langle n \rangle)}$, can be used to derive the
result obtained in \cite{elf} for $\beta_{\mathrm{min}}$
(Supplementary Information).  Our analysis indicates that this rate
corresponds to $\beta_{\mathrm{min}} = P_{0}k_{0}{b}/{(b+1)}$
which implies that for noise reduction, the rate of TF dissociation
must be greater than the rate of arrival of nonzero bursts of proteins.

Our results can also be used to analyze the optimal value
$\beta=\beta_c$ at which noise suppression is maximal.  The results
derived in \cite{elf} using moment-closure approaches serve as a good
approximation in the limit of large $\langle n \rangle$.  Since our
exact results apply for arbitrary parameter values, they can be used
to connect large $\langle n \rangle$ results with those for low
$\langle n \rangle$.  As discussed below, this analysis leads to some
interesting observations.

\begin{figure}[h]
 \centering
 \includegraphics[height=4cm,width=5.5cm]{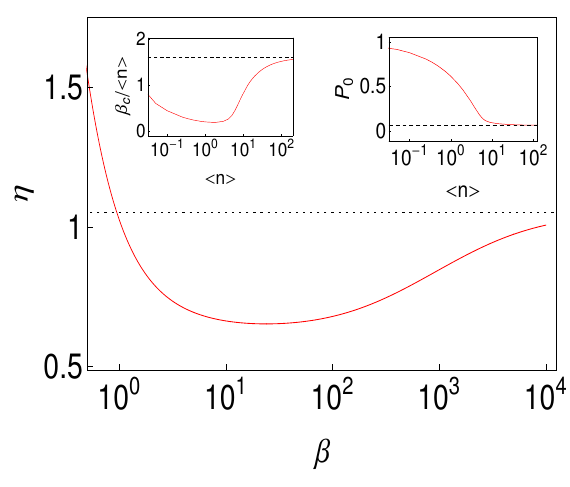}
 \caption{Optimization of noise suppression in negative feedback:
   Noise is shown as a function of dissociation rate $\beta$ for
   $\langle n \rangle=b=20$. Corresponding variations for the optimal
   dissociation rate $\beta_c$ and the probability $P_0$ are plotted
   for different values of $\langle n \rangle$ in the inset, dotted
   lines representing the prediction of Ref. \cite{elf}. Other
   parameters are: $\mu=1,\alpha=0,\tilde{\alpha}=25$.}
 \end{figure}

In Ref. \cite{elf}, it was shown that the optimal value of
dissociation rate $\beta_c$ is linearly dependent on $\langle n
\rangle$, i.e. $\beta_c/\langle n \rangle$ remains constant as
$\langle n \rangle$ is varied by changing $k_0$. As expected, we
recover this feature when $\langle n \rangle$ is large (see
Fig. 3). However, for small $\langle n \rangle$ we see a strong
deviation from the large $\langle n \rangle$ limit, characterized by
non-monotonic variation. Furthermore, this non-trivial variation in
the optimal dissociation rate is also reflected in the probability of
the promoter state being transcriptionally active, $P_0$.  As shown in
the figure, $P_0$ decreases monotonically with $\langle n \rangle$ and
in the limit of large $\langle n\rangle$ it approaches the result
derived in Ref. \cite{elf}. Thus our results predict that, for optimal
noise suppression in low abundance TFs, the fraction of time that the
promoter is active decreases as we increase TF concentration.

\noindent{\bf Switching statistics:} As noted in Fig. 1B, when
$k_0=k_1$, the model analyzed can be mapped to a two-state system
driven by a \emph{bursty} input signal. Several cellular systems can
be modeled (at a coarse-grained level) as two-state switches; thus it
is of interest to explore how such switches respond to fluctuating
inputs~\cite{koshland1976,berg1977,cluzel11,levin}.  The results
obtained in this work lead to exact analytical expressions for the
corresponding switch statistics.

The quantity of interest is the variance of the switch,
$\sigma^2=P_0(1-P_0)$, with $P_0$ given by Eq.(\ref{eP0}). Note that
Eq.(\ref{eP0}) is valid for proteins produced in geometrically
distributed bursts with mean burst size $b$. On the other hand,
previous work \cite{levin} has considered the case such that each
burst leads to creation of exactly one protein (i.e. protein dynamics
is a simple birth-death process). Remarkably, there exists a mapping
between the analytical solutions in these two cases (Supplementary
Information).  Using this mapping, we obtain the following exact
result for the problem considered in previous work \cite{levin}
\begin{eqnarray}{\label{eP02}}
P_0&=&\frac{\beta(\mu+\tilde{\alpha})}{\tilde{\alpha}k+(\alpha+\beta)(\mu+\tilde{\alpha})}\\
&\times& _1F_1\left[1,1+\frac{\alpha+\beta}{\mu+\tilde{\alpha}}
+\frac{\tilde{\alpha}k}{(\mu+\tilde{\alpha})^2},-\frac{k}{\mu}\left(\frac{\tilde{\alpha}}{\mu+\tilde{\alpha}}\right)^2\right].\nonumber
\end{eqnarray}

As expected, the above expression, Eq.(\ref{eP02}), reduces to
analytical results derived in \cite{levin} (for $\alpha=0$) in
limiting cases.  For example, in the slow switching limit,
i.e. $\tilde{\alpha}k\ll \mu=1$, Eq. (\ref{eP02}) leads to
$P_1=1-P_0=(\tilde{\alpha}/(1+\tilde{\alpha}))k/(\beta+(\tilde{\alpha}/(1+\tilde{\alpha})k)$,
which is identical to the result obtained in \cite{levin}. Similarly,
in the fast switching limit $\tilde{\alpha}k\gg \mu=1$, if we further
set $\tilde{\alpha}\rightarrow \infty$ and $k\ll\mu=1$, we obtain
$P_1=k(1+\beta)/(k+\beta)$, consistent with the result obtained in
\cite{levin}.  The exact result derived above, Eq.(\ref{eP02}),
allows for analysis of switching statistics beyond these limits, i.e.
throughout parameter space.

\begin{figure}
\centering
\includegraphics[bb=1 45 431 250, height=4cm,width=8cm]{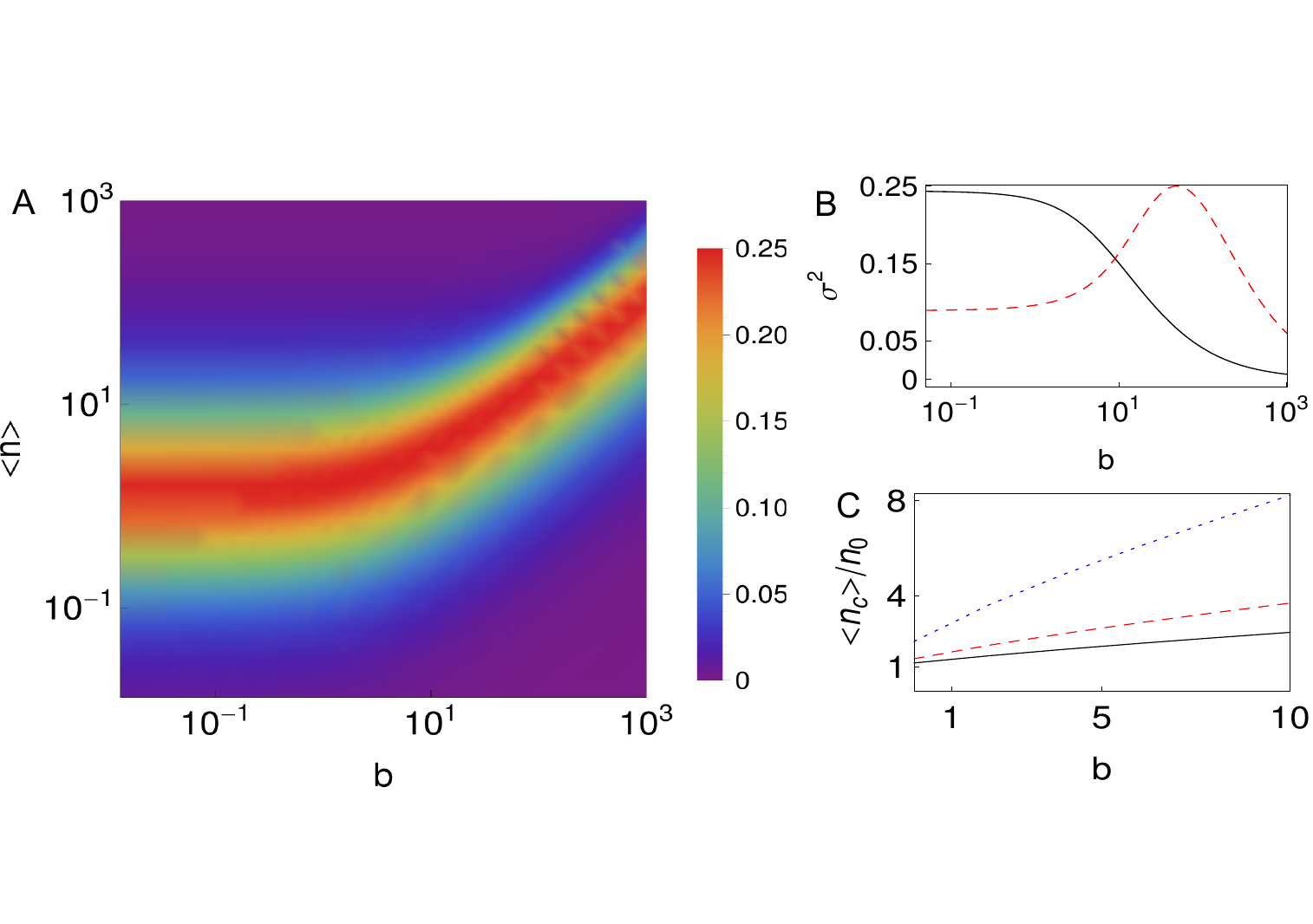}
\caption{Two-state switch statistics. A) Density plot for $\sigma^2$
  is shown as a function of $b$ and $\langle n\rangle$.  B) Variations
  of $\sigma^2$ with $b$ for two different values of $\langle
  n\rangle$, 1(solid) and 10(dashed). In both A and B,
  $\tilde{\alpha}=1$.  C)Variation of $\langle n_c \rangle/n_0$ with
  $b$, different lines correspond to different values of feedback
  strength $\tilde{\alpha}=0.5$ (solid), $1$ (dashed), $3$
  (dotted). In all plots, other parameters are: $\Delta k=\alpha=0$,
  $\beta=\mu=1$.}
\label{fig:switch}
\end{figure}

Furthermore, the results derived can be used to explore how bursty
protein production affects switching statistics.
Fig.\ref{fig:switch}A shows how the switch variance $\sigma^2$ depends
on the burst size, $b$, and the average number of proteins, $\langle n
\rangle$.  Some interesting observations can be made which
highlight the nontrivial variation of $\sigma^2$ with bursting. For
large $\langle n \rangle$ values, $\sigma^2$ shows a non-monotonic
variation with $b$, with a maximum at a critical burst size, $b_c$
(see Fig.\ref{fig:switch}B). On the other hand, for low $\langle n
\rangle$, we observe that $\sigma^2$ decreases monotonically with $b$
with the maximum corresponding to $b_c\rightarrow 0$
(Fig.\ref{fig:switch}B). These different behaviors can be understood based on the following observations: 1) For fixed $\langle n \rangle$, 
$P_0$  increases with increasing burst size $b$, and 2), for fixed $b$,  $P_0$ decreases with increasing $\langle n \rangle$.
Thus, in the limit $b\rightarrow 0$, for $\langle n \rangle$ such that $P_0\ge1/2$ we obtain a monotonic decrease in the 
variance $\sigma^2=P_0(1-P_0)$ as $b$ is increased. On the other hand, for $\langle n \rangle$ such 
that  $P_0<1/2$ (in the limit $b \rightarrow 0$) we obtain non-monotonic variation with $b$.

Next, we focus on the variation of $\sigma^2$ with mean protein
$\langle n \rangle$ for a fixed $b$. As can be seen in
Fig.\ref{fig:switch}A, it shows a non-monotonic variation, with
$\sigma^2$ being maximum at a critical mean protein level, $\langle
n_c \rangle$. Often, it is of interest to estimate the value $\langle
n_c \rangle$ which maximizes the noise in switching statistics. In
Fig.\ref{fig:switch}C, we compare the mean-field estimate $n_0$
(obtained by replacing the fluctuating $n$ by its mean value, $\langle
n \rangle$) with the corresponding exact value.  As indicated in the
figure, deviation from the mean-field estimate is significant and
increases with increasing burst-size and feedback strength.  The
analytical results derived are thus useful in obtaining accurate
estimates of parameters that maximize noise in switching statistics.

To conclude, we have studied an exactly solvable model that integrates
key features of regulation of gene expression, specifically: bursting,
promoter switching and feedback, in a single model.  
The derived results provide new insights into the roles of bursting
and feedback (both positive and negative) in fine-tuning noise in
protein distributions. Furthermore, the results obtained can serve as
building blocks for future studies focusing on noise optimization
strategies by varying the underlying parameters.  The model developed
can also be applied to study the statistics of a simple two-state
switch driven by a bursty protein noise source. Our results show
that such bursty input noise can induce strong deviations in the
optimal parameters for switch variance from the corresponding
mean-field predictions. The development of analytical approaches, as
outlined in this work, is thus an important ingredient for accurate
quantitative modeling of stochastic cellular processes.

The authors acknowledge funding support from the NSF through Awards
No. PHY-1307067 and DMS-1413111.

\pagebreak
\widetext
\begin{center}
\textbf{\large Supplementary Material:}
\end{center}
\setcounter{equation}{0}
\setcounter{figure}{0}
\setcounter{table}{0}
\makeatletter
\renewcommand{\theequation}{S\arabic{equation}}
\renewcommand{\thefigure}{S\arabic{figure}}
\renewcommand{\bibnumfmt}[1]{[S#1]}
\renewcommand{\citenumfont}[1]{S#1}
\section{Solution for the steady state protein distribution}
In the steady state, Eq. (2) in the main text can be written as
\begin{eqnarray}{\label{ed1}}
\beta G_1&=&\left[k_0(1-\tilde{g})+\alpha\right]G_0-\left[\mu(1-z)-\tilde{\alpha}z\right]\partial_zG_0,\nonumber \\
\alpha G_0+\tilde{\alpha}z\partial_zG_0&=&\left[k_1(1-\tilde{g})+\beta\right]G_1-\mu(1-z)\partial_zG_1.
\end{eqnarray}
By eliminating $G_1$ in Eq. (\ref{ed1}), we can write a single equation in terms of $G_0$,
 \begin{eqnarray}{\label{ed2}}
A(z)G_0(z)-B(z)\partial_zG_0(z)+C(z)\partial_z^2G_0(z)=0,
\end{eqnarray}
with
\begin{eqnarray}{\label{ed3}}
A(z)&=&(1-\tilde{g})\left[ k_0k_1(1-\tilde{g})+\alpha k_1+\beta k_0\right]+\mu (1-z)k_0\partial_z \tilde{g}, \nonumber\\
B(z)&=&\mu(1-z)\left[ (k_0+ k_1)(1-\tilde{g})+\alpha+\tilde{\alpha}+\beta+\mu \right]-z\tilde{\alpha}k_1(1-\tilde{g}),\nonumber \\
C(z)&=&\mu(1-z)\left[\mu(1-z)-z\tilde{\alpha}\right].
\end{eqnarray}
Writing $G_1=G-G_0$ in the second equation of Eq. (\ref{ed1}), we can express $G_0$ in terms of $G$ which reads as
\begin{eqnarray}{\label{ed4}}
(k_1-k_0)(1-\tilde{g})G_0=k_1(1-\tilde{g})G-\mu(1-z)\partial_zG.
\end{eqnarray}
Using Eq. (\ref{ed4}) in the first equation of Eq.(\ref{ed1}), and taking 
\begin{equation}{\label{ed5}}
\tilde{g}=\frac{1}{1+b(1-z)}
\end{equation}
followed by a transformation $x=1-z$ leads to the following equation for $G$.
\begin{eqnarray}{\label{ed6}}
A(x)G(1-x)+B(x)\partial_xG(1-x)+C(x)\partial_x^2G(1-x)=0,
\end{eqnarray}
where
\begin{eqnarray}{\label{ed7}}
A(x)&=&\frac{b}{\left[1+bx\right]^2}\left[ bx\left\{k_0k_1+\alpha k_0+\beta k_1 \right\}+ \alpha k_0+\beta k_1\right], \nonumber \\
B(x)&=& \frac{\mu}{1+bx}\left[ bx\left\{ k_0+k_1+\alpha+\beta+\mu\right\} + \alpha+\beta \right] - \frac{\tilde{\alpha}(1-x)}{1+bx}b(k_1+\mu),\nonumber \\
C(x)&=&\mu\left[x(\mu+\tilde{\alpha}) - \tilde{\alpha} \right].
\end{eqnarray}
Applying the transformation, $G(1-x)=\exp[f(x)]F(x)$, Eq. (\ref{ed6}) can be rewritten as
\begin{eqnarray}{\label{ed8}}
\AA(x)F(x)+\BB(x)\partial_xF(x)+\CC(x)\partial_x^2F(x)=0,
\end{eqnarray}
with 
\begin{eqnarray}{\label{ed9}}
\AA(x)&=&A(x)+B(x)\partial_xf+C(x)\left[\partial_x^2f+(\partial_xf)^2\right],\nonumber\\
\BB(x)&=&B(x)+2C(x)\partial_xf,\nonumber\\
\CC(x)&=&C(x).
\end{eqnarray}
Setting $f(x)=-(k_1/\mu)\ln(1+bx)$ and changing $x$ to $\xi$ using the transformation, $\xi=\phi(1+ bx)$
with $\phi=(\mu+\tilde{\alpha})/(\mu+\tilde{\alpha}(1+b))$,
the resulting equation for $F(x)$ reduces to the hypergeometric differential equation:
\begin{equation}{\label{hyperEq}}
uvF(\xi)+[(u+v+1)\xi-w]\partial_\xi F(\xi)+\xi(\xi-1)\partial_\xi^2 F(\xi)=0,
\end{equation}
where $u$, $v$ and $w$ are given by Eq.(4), and corresponding solution for $G(z)$ is given in Eq. (3) in 
the main text.

\section{Limiting cases of the exact generating function}
 Here we discuss how the exact steady-state generating function,  $G(z)$, reduces to known results 
 in different limiting cases: \\
(1) In the absence of feedback, i.e. $\tilde{\alpha}=0$, the generating function reduces to
 \begin{equation}
 G(z)={_2F_1}[u,v|\alpha+\beta|b(z-1)]/[1+b(1-z)]^{k_1/\mu},
 \end{equation}
 which is identical to the result first derived in \cite{swain08}.  \\
(2) When the protein production rate in the inactive state vanishes 
 ($k_0=0$) and for no spontaneous switching to the active state ($\alpha=0$), the system reaches, in the long-time limit, an absorbing state with no proteins.
 Correspondingly, the expression for the generating function reduces to
 $G(z)=1$. \\
(3) In the limit $\alpha=\tilde{\alpha}=0$, we have a model
 with constitutive production of geometric bursts with rate
 $k_0$. It follows that 
 \begin{equation}
 G(z)=1/[1+b(1-z)]^{k_0/\mu},
 \end{equation}
 which is the
 generating function for a negative binomial distribution as expected
 \cite{swain08}. \\ 
 (4) For $\beta=0$ or $\Delta k=0$, we get
 \begin{equation}
  G(z)=1/[1+b(1-z)]^{k_1/\mu}.
 \end{equation}
 Thus the result obtained encompasses previously 
 derived results and extends them to include the effects of bursting and feedback.

\section{Deriving the steady-state probability of promoter being in state $0$: $P_0$}

Starting from (2), and adding the equations for $G_0$ and $G_1$, we get
\begin{equation}
G_0=\frac{k_1}{k_1-k_0} G-\frac{\mu}{k_1-k_0}\frac{(1-z)}{(1-\tilde g)}\partial_zG.
\end{equation}
Taking the limit $z=1$, this leads to
\begin{equation}
P_0=\frac{k_1}{k_1-k_0}-\frac{\mu}{k_1-k_0}\frac{\langle n\rangle}{b}.
\end{equation}
Replacing $\langle n\rangle$ by its value we get
\begin{equation}
P_0=\frac{\mu}{\Delta k}\phi\frac{uv}{u+v+1-w}\frac{{_2F_1}[u+1,v+1,u+v+2-w|1-\phi]}{{_2F_1}[u,v,u+v+1-w|1-\phi]},
\end{equation}
which can be expressed as equation (5).

\section{Expressions for $\langle n \rangle$ and $\alpha_{\text{eff}}$}

Here we derive expressions for the mean protein levels and the constant rate of switching from the state 0 to state 1
in the effective model that maintains the same mean protein numbers as that of the original model. 
From the expression for the steady-state generating function, we obtain the following expression for the mean number of proteins   
\begin{eqnarray}{\label{eavn}}
\langle n\rangle/b=\frac{k_1}{\mu}+\phi \frac{uv}{u+v+1-w}\frac{_2F_1[u+1,v+1|u+v+2-w|1-\phi]}{_2F_1[u,v,u+v+1-w|1-\phi]}.
\end{eqnarray}
For the effective model with a constant rate of promoter transitions we have
\begin{equation}
 \langle n \rangle/b=\frac{k_0}{\mu}\left(\frac{\beta}{\alpha_{\text{eff}}+\beta}\right)+\frac{k_1}{\mu}
\left(\frac{\alpha_{\text{eff}}}{\alpha_{\text{eff}}+\beta}\right).
\end{equation}
Equating the two means, we get the general equation determining $\alpha_{\text{eff}}$.
The specific cases of positive and negative feedback in the main text 
are discussed below.

For positive feedback (with $k_0$=0), the mean number of proteins for the original model can be expressed as
 \begin{equation}{\label{esn1}}
  \langle n \rangle=P_1\frac{k_1b}{\mu},
 \end{equation}
whereas for 
the effective model with the same mean we have
\begin{equation}{\label{esn2}}
\langle n \rangle=\left(\frac{\alpha_{\text{eff}}}{\alpha_{\text{eff}}+\beta}\right)\frac{k_1b}{\mu}.
\end{equation}
Using Eqs.(\ref{esn1}) and (\ref{esn2}) we get 
\begin{equation}
 \alpha_{\text{eff}}= \frac{\langle n \rangle\mu\beta}{k_1b-\langle n\rangle\mu}.
\end{equation}
Similarly, for negative feedback (with $k_1=0$) we get
\begin{equation}
 \alpha_{\text{eff}}=\frac{k_0b-\langle n\rangle\mu}{\langle n\rangle\mu}\beta.
\end{equation}

\section{Mapping from model with arrival of geometric bursts to model with Poisson arrivals}

Here we illustrate how we can map a model with arrival of proteins in geometric bursts 
to a model for which each arrival leads to the production of a single protein.

First, consider the geometric burst distribution conditional on the
production of at least one protein (i.e. ignore bursts which do not
result in the production of any proteins).  If the original burst
distribution has mean $b$, then the generating function of the
corresponding conditional geometric distribution is given by $g'(z) =
\frac{z}{1 + b(z-1)}$.  It is straightforward to see that the mean of
the conditional geometric distribution is $b' = 1+b$. If we now take
the limit $ b \rightarrow 0$ for the conditional geometric
distribution, this corresponds to the case in which exactly one
protein is produced in every burst. Thus the simple Poisson arrival
process can be recovered as a limit of the model with arrival of {\em
  conditional} geometric bursts.

Now we note that the model described in Fig. 1B can be equivalently
viewed from two perspectives: (i) geometric bursts (with mean $b$)
arriving at rate $k$ or (ii) conditional geometric bursts (with mean
$b'=1+b$) arriving with rate $k'= kb/(1+b)$, where $b/(1+b)$ is the
probability that the burst produces at least one protein.  Thus the
steady-state solution obtained in the main text is also the solution
for a model with arrival of conditional geometric bursts. Based on
this observation, the results obtained also lead to exact results for
models with conditional geometric bursts, using the mapping
$$b'= 1+b ~~\text{and}~~ k'=kb/(1+b).$$ Carrying out this mapping and
taking the limit $b\rightarrow 0$ in Eq. (5) leads to the
exact result for the problem considered in previous work \cite{levin}.

\section{Deriving the minimum dissociation rate: $\beta_{\text{min}}$}
Using Eq. (9), expression for the mean number of proteins is
\begin{equation}{\label{eb1}}
 \langle n \rangle=\frac{buv}{u+v+1-w}\phi\lambda,
\end{equation}
where
\begin{equation}{\label{eb2}}
 \lambda=\frac{_2F_1[u+1,v+1,c+1,1-\phi]}{_2F_1[u,v,c,1-\phi]},
\end{equation}
$c=u+v+1-w$ and  expressions for $u$, $v$, $w$ and $\phi$ are given by
Eq. (4). Similarly, using Eq. (3), the exact expression for the protein noise can be 
written as a sum of two terms
\begin{equation}{\label{eb3}}
 \eta=\eta_{\infty}+\eta_{\tilde{\alpha}},
\end{equation}
where $\eta_{\infty}=\frac{1+b}{\langle n \rangle}$
is the noise contribution when proteins are produced constitutively, i.e. in the limit of $\beta\rightarrow \infty$,
and 
\begin{equation}{\label{eb4}}
\eta_{\tilde{\alpha}}=\delta+(1+\delta)\frac{b\phi\lambda}{\langle n\rangle c}
\left[c\left(1-\frac{1}{\phi\lambda(1+\delta)}\right)+\frac{1+u+v-c}{c}-\frac{1+u+v+uv}{c(c+1)}\right]
\end{equation}
is the noise contribution due to promoter switching, and $\delta$ is given as
\begin{equation}{\label{eb5}}
 \delta=\frac{ _2F_1[u,v,c,1-\phi] _2F_1[u+2,v+2,c+2,1-\phi]}{ _2F_1[u+1,v+1,c+1,1-\phi]^2}-1.
\end{equation}

It is interesting to note that when the transition rate from the state
0 to 1 is a constant, the second term $\eta_{\tilde{\alpha}}$ is a
positive definite quantity and leads to monotonically increasing
noise with decreasing $\beta$. However, in the presence of feedback, 
the second term can be positive or negative or zero. The
value of finite $\beta$ where $\eta_{\tilde{\alpha}}=0$ corresponds to
$\beta_{\text{min}}$, which we wish to find.  This is the minimum
value of $\beta$ which marks the beginning of noise suppression
due to negative feedback.

Using the approximation $\delta\ll 1$ leads to the
equation that the minimum dissociation rate has to satisfy
\begin{equation}{\label{eb6}}
 \left(\frac{1+u+v-c}{c}-\frac{1+u+v+uv}{c(c+1)}\right)+\left(1-\frac{1}{\phi\lambda}\right)=0,
\end{equation}
i.e. at $\beta_{\text{min}}$, the sum of both bracketed terms should
vanish. 

To proceed further, we need to determine how $k_0$ varies
with $\beta$, given the constraint that $\langle n \rangle$ is held
fixed. This can be simplified by making the 'mean-field'
approximation:
\begin{equation}
P_0=\frac{\beta}{\beta+\tilde{\alpha}\langle n \rangle}
\end{equation}
in combination with the exact equation $\langle n
\rangle=P_0k_0b$. Defining $x = \frac{\langle n \rangle}{\beta}$, the
preceding approximation gives us $k_0 = \frac{\langle n \rangle}{b} (1
+ \tilde{\alpha} x)$. Substituting this in Eq. (\ref{eb6}), we obtain the solution
\begin{equation}{\label{eb8}}
 \beta_{\text{min}}=\frac{\langle n \rangle}{1+b}.
\end{equation}
This is identical to the result derived in \cite{elf}. The preceding
analysis highlights the assumptions needed to obtain this result from
the exact results derived in the main text. The derivation also shows that 
we have 
\begin{equation}
\beta_{\text{min}} = P_0\frac{k_0b}{1+b}. 
\end{equation}
The right-hand side of the above equation represents the rate of arrival of non-zero bursts of proteins 
at steady-state.  Thus the minimum dissociation rate has to be greater than the rate of arrival 
of non-zero bursts of protein production.

\end{document}